\documentclass[11pt,a4paper]{article}
\usepackage{t1enc}
\usepackage[latin1]{inputenc}
\usepackage[english]{babel}
\normalfont
\usepackage{amsmath}
\usepackage{yfonts}
\usepackage{bbm}
\usepackage{bm}
\usepackage{wasysym}
\usepackage{amssymb}
\usepackage{mathrsfs}
\usepackage{marvosym}
\usepackage{pifont}
\usepackage{hyperref}

\newcommand{\be}[0]{\begin{equation}}
\newcommand{\ee}[0]{\end{equation}}
\newcommand{\tcA}[0]{\tilde{\mathcal{A}}}
\newcommand{\cA}[0]{\mathcal{A}}
\newcommand{\cG}[0]{\mathcal{A}}

\newcommand{\tcC}[0]{\tilde{\mathcal{C}}}
\newcommand{\cCn}[0]{\mathcal{C}_n}
\newcommand{\cC}[0]{\mathcal{C}}
\newcommand{\cTn}[0]{\mathcal{T}_n}

\newcommand{\tsH}[0]{\tilde{\mathscr{H}}}
\newcommand{\sH}[0]{\mathscr{H}}

\numberwithin{equation}{section}

\begin{document}

\vspace*{-1cm}
\thispagestyle{empty}
\begin{flushright}
LPTENS 10/46 
\end{flushright}
\vspace*{1.5cm}

\begin{center}
{\Large 
{\bf Geometric quantization and\vspace{.5cm}\\the metric dependence of the self-dual field theory}}
\vspace{2.0cm}

{\large Samuel Monnier}%
\vspace*{0.5cm}

Laboratoire de Physique Th\'eorique de l'\'Ecole 
Normale Sup\'erieure
\\
CNRS UMR 8549, \\
24 rue Lhomond, 75231 Paris Cedex 05, France \\
monnier@lpt.ens.fr
\\ 

\vspace*{2cm}

{\bf Abstract}
\end{center}

We investigate the metric dependence of the partition function of the self-dual $p$-form gauge field on an arbitrary Riemannian manifold. Using geometric quantization of the space of middle-dimensional forms, we derive a projectively flat connection on its space of polarizations. This connection governs metric dependence of the partition function of the self-dual field. We show that the dependence is essentially given by the Cheeger half-torsion of the underlying manifold. We compute the local gravitational anomaly and show how our derivation relates to the classical computation based on index theory. As an application, we show that the one-loop determinant of the (2,0) multiplet on a Calabi-Yau threefold coincides with the square root of the one-loop determinant of the B-model.

\newpage

\tableofcontents

\section{Introduction and summary}

The self-dual gauge field is a $p$-form gauge field whose field strength is constrained by a self-duality condition. In a spacetime with Lorentz signature, real self-dual gauge fields can exist only in dimension $4\ell + 2$. They appear on the two-dimensional worldsheet of the heterotic string, on the six-dimensional worldvolume of the M5-branes and type IIA NS5-branes as well as in the ten-dimensional spacetime of type IIB supergravity. Yet, this theory has remained mysterious and ill-understood for a long time because of the absence of a simple covariant Lagrangian description. 

Witten proposed in \cite{Witten:1996hc} a radically new way of studying the self-dual field theory. He argued that the partition function of the self-dual $2\ell$-form gauge field theory on a $4\ell + 2$ manifold $M$, as a function of an external $2\ell +1$-form gauge field, could be constructed by geometric quantization of the space $\tilde{\mathcal{A}}$ of $2\ell + 1$-forms on $M$. $\tilde{\mathcal{A}}$ carries a natural symplectic structure, given by the intersection product, which is antisymmetric in this degree.

Let us recall the basics of K\"ahler geometric quantization \cite{MR1183739, Axelrod:1989xt} and see how it can be used to understand the metric dependence of the partition function. Given a symplectic manifold $\tilde{\mathcal{A}}$, one first pick a complex structure that turns the symplectic form into a K\"ahler form. One then has to construct a ``prequantum'' holomorphic line bundle whose first Chern class coincides with the cohomology class of the symplectic form. The ``quantum'' Hilbert space associated with the quantum system is defined as the space of holomorphic sections of the prequantum bundle. This procedure involves an arbitrary choice of a complex structure on $\tilde{\mathcal{A}}$. In order for the quantization to be independent of this choice, one has to provide a way of identifying the Hilbert spaces obtained with different complex structures. This is performed by considering the family of Hilbert spaces constructed in this way as a bundle over the space of complex structures, and by providing a projectively flat connection on this bundle. The latter allows to identify canonically rays of vectors, i.e. quantum states, in neighboring fibers.

In the case of interest to us, up to identifications given by large gauge transformations, the symplectic space $\tilde{\mathcal{A}}$ is an infinite-dimensional affine space. Choosing a metric on the manifold $M$ on which the theory is defined naturally endows $\tilde{\mathcal{A}}$ with a complex structure, given by the Hodge star operator, which squares to $-\mathbbm{1}$ on $\tilde{\mathcal{A}}$. Moreover, the quantum Hilbert space is one-dimensional and according to Witten's argument, the unique section of the prequantum bundle is the partition function of the self-dual field, as a function of the external gauge field. We see therefore that the projectively flat connection provided by geometric quantization, which relates sections of the prequantum bundle for different choices of complex structures, provides a mean to understand the metric dependence of the partition function of the self-dual field theory.

In this work, we carry out this program in detail. We will see that the projectively flat connection on the quantum bundle does indeed determines the metric dependence of the partition function, albeit not in a form as explicit as one might have wished. The connection contains a central term given by a certain product of regularized determinants of Laplacians, very reminiscent of Ray and Singer's analytic torsion \cite{MR0295381,RaySinger1973,MR0339293}. This object (or more precisely its square) appeared at least once in the mathematical literature under the name of Cheeger's half-torsion \cite{Branson2005}. It does not enjoy the nice topological invariance properties of the Ray-Singer torsion, and even its dependence on conformal transformations of the metric seems very involved \cite{Branson2005}. The square root of the Cheeger half-torsion can essentially be seen as the norm of the one-loop determinant of the self-dual field. If it is set aside, a simple dependence on the metric remains, factoring through the restriction of the Hodge star operator on harmonic forms and captured by a Siegel theta function \cite{Witten:1996hc}. Therefore all the non-triviality of the metric dependence is contained in Cheeger's half-torsion.
In an appendix to this paper, we specialize to the dimension six case that is of interest for the five-branes in type IIA string theory and M-theory. Using our result about the self-dual field, we compute the norm of the one-loop determinant of the (2,0) multiplet on a Calabi-Yau threefold, and show that it coincides with the square root of the Bershadsky-Cecotti-Ooguri-Vafa (BCOV) torsion \cite{Bershadsky:1993cx, fang-2008-80}. The BCOV torsion has a much simpler dependence on the metric of the underlying Calabi-Yau: as it can be expressed in terms of complex Ray-Singer torsions, it is independent of the K\"ahler class of the metric. The BCOV torsion arises as the norm of the one-loop determinant of the B-model, so this result confirms the link between the five-brane worldvolume theory and the B-model conjectured in \cite{Dijkgraaf:2002ac, Alexandrov:2010ca}. We should however stress that the determinant of the (2,0) multiplet is really a square root of the determinant of the B-model. Moreover, as our argument makes a crucial use of the covariantly constant spinors existing on Calabi-Yau manifolds, it seems unlikely that this equality would continue to hold on more general (complex) six-manifolds.

The non-trivial curvature of the connection on the quantum bundle reflects the fact that the self-dual field displays a local gravitational anomaly. We show how Cheeger's half-torsion can be seen as the torsion of a certain complex. By folding this complex, we recover the Dirac operator appearing in the familiar derivation of the local anomaly using index theory \cite{AlvarezGaume:1983ig}. Sadly, geometric quantization says nothing about the global anomaly (the holonomies of the connection). The reason for this is simple: geometric quantization yields a connection on the space of polarizations, or complex structures, of the symplectic space to be quantized. In our case, this is the space of Hodge star operators modulo diffeomorphisms isotopic to the identity. Global anomalies are associated to ``large'' diffeomorphisms that are not isotopic to the identity (equivalently to elements of the mapping class group of $M$) and therefore cannot be obtained from geometric quantization. Put differently, geometric quantization only produces a connection on the universal covering of the space of metrics modulo diffeomorphisms, and there is no canonical way to push it down to a connection on the space of metrics modulo diffeomorphisms. The computation of the gravitational anomaly of the self-dual field should really be addressed, as the only available formula \cite{Witten:1985xe} is valid only in the case where the manifold $M$ has vanishing middle-dimensional cohomolgy. A knowledge of the global gravitational anomaly would be necessary to check the cancellation of global gravitational anomalies in M-theory backgrounds including M5-branes (see for instance \cite{Henningson:1997da}). We hope to show in a future paper that the work of Hopkins and Singer \cite{hopkins-2005-70} allows to settle this question. 

Here is a summary of the organization of the paper. Sections \ref{SecGeomQuant} and \ref{SecSDF} are devoted to the construction of a projectively flat connection on the quantum bundle. 
Section \ref{SecGeomQuant} is an introduction to the method of geometric quantization, applied to affine spaces. In section \ref{SecSDF}, we apply the results of section \ref{SecGeomQuant} to the case of the self-dual field. We derive explicitly the projectively flat connection governing the metric dependence of the partition function. In section \ref{SecAnom}, we compute the local anomaly. We show in section \ref{SecMetricDep} that the anomaly affects only the phase of the partition function, and that its well-defined norm is essentially given by the square root of Cheeger's half-torsion. We make contact with the usual derivation of the local anomaly through index theory in section \ref{SecIndex}. Finally, in appendix \ref{SecOne-Loop}, we compute the one-loop determinant of the (2,0) supermultiplet on a Calabi-Yau threefold.

\section{Geometric quantization}

\label{SecGeomQuant}

In this section, we review the K\"ahler geometric quantization of affine spaces. Although this is standard material (see \cite{Axelrod:1989xt}), we pay a particular attention to the measure on the reduced phase space induced by symplectic reduction. This point is crucial to derive the correct projectively flat connection on the quantum bundle and obtain the dependence of the partition function of the self-dual field on the metric.

\subsection{Geometric quantization of affine spaces}
 
Suppose we are given a real affine space $\mathcal{A}$ of dimension $2n$ endowed with a symplectic form $\omega$. 

Let $J$ be a complex structure on $\mathcal{A}$ compatible with the affine structure of $\mathcal{A}$. We furthermore require that $\omega(.,J.)$ is positive-definite and that $\omega$ is of type $(1,1)$. Such a complex structure turns $\mathcal{A}$ into a K\"ahler manifold. We choose global holomorphic and antiholomorphic coordinates $\{z^i\}$ and $\{\bar{z}^i\}$ on $\mathcal{A}$. We will find it useful to follow the convention of \cite{Axelrod:1989xt} and underline holomorphic indices (e.g. $\underline{i}$) and overline antiholomorphic ones (e.g. $\overline{i}$).

An affine complex structure on $\mathcal{A}$ can be characterized by its $\pm i$ eigenspaces on the complexification of the tangent bundle of $\mathcal{A}$: $T_{\mathbbm{C}}\mathcal{A} = \mathcal{W} \oplus \bar{\mathcal{W}}$. The latter is actually a positive polarization of $T_{\mathbbm{C}}\mathcal{A}$, namely a decomposition into maximally isotropic subspaces such that $-i\omega(v_1, \bar{v}_2) > 0$ for all nonzero $v_1, v_2 \in W$. The latter condition follows from the fact that $\omega(.,J.)$ is positive definite. A reference positive polarization $\mathcal{W}_0$ is mapped onto another positive polarization under any real symplectomorphism of $T_{\mathbbm{C}}\mathcal{A}$. This action turns out to be transitive, with stabilizer given by the group of unitary transformations of $\mathcal{W}$. The set of affine complex structures can therefore be identified with the symmetric space ${\rm Sp}(2n, \mathbbm{R})/U(g)$. This coset space has an alternative description as the Siegel upper half-plane, namely the set of symmetric $n \times n$ matrices with positive definite imaginary part. We will sometimes refer to the space of complex structures as the period domain, and write it $\cCn$. Note that $\cCn$ is contractible.

Let $\mathscr{L}$ be a prequantum bundle, i.e. a line bundle on $\mathcal{A}$ equipped with a connection $\nabla$ whose curvature equals $-i\omega$. As $\omega$ is of type $(1,1)$, we have $[\nabla_{\overline{i}}, \nabla_{\overline{j}}] = 0$ and $\mathscr{L}$ is automatically equipped with a holomorphic structure. The prequantum bundle can be identified with the trivial line bundle over $\mathcal{A}$ with Hermitian structure $|\psi|^2 = \exp(-h) \bar{\psi}\psi$, where $h$ is a complex function such that $\bar{\partial}\partial h = -i\omega$.

The Hilbert space $\mathscr{H}|_J$ associated to the quantization of $\mathcal{A}$ is given by the holomorphic square integrable sections of $\mathscr{L}$. The phase space coordinates $z^i$ are quantized into multiplication operators by $z^i$, while the coordinates $\bar{z}^i$ act as $\frac{\partial}{\partial z^i}$.

As the notation suggests, $\mathscr{H}|_J$ depends on the complex structure $J$, which was not part of the original quantization problem. As it stands, the quantization procedure yields a quantum bundle over the space $\cCn$ of affine complex structures, with fiber $\mathscr{H}|_J$ over $J \in \cCn$. In order to insure the independence of the quantization on the choice of $J$, we have to construct a projectively flat connection on the quantum bundle. Such a connection allows to identify canonically the rays of $\mathscr{H}|_J$ with those of $\mathscr{H}|_{J+\delta J}$.

Note that $\cCn$ is naturally a complex manifold because its tangent space decomposes into infinitesimal deformations of holomorphic and antiholomorphic functions on $\mathcal{A}$. Let $\delta$ be the differential on $\cCn$. Its holomorphic and antiholomorphic components are denoted by $\delta^{(1,0)}$ and $\delta^{(0,1)}$, respectively. $\delta J$ is a 1-form on $\cCn$. 

In the case when $\mathcal{A}$ is a finite dimensional affine space, the connection $\delta_{\mathscr{H}}$ on $\mathscr{H}$ has a simple expression (\cite{Axelrod:1989xt}, section 1):
\begin{equation}
\label{EqProjConn}
\delta_{\mathscr{H}} = \delta - Q \qquad Q = -\frac{1}{4}(\delta J \omega^{-1})^{\underline{i} \underline{j}} \nabla_{\underline{i}} \nabla_{\underline{j}}  \;.
\end{equation}
This connection has the following crucial properties:
\begin{itemize}
	\item It preserves holomorphicity, i.e. it maps holomorphic sections of the prequantum bundle to holomorphic sections. This is a necessary condition to ensure that the connection preserves the fibers of the quantum bundle.
	\item It is projectively flat, so that rays of holomorphic sections of the quantum bundle are canonically identified along $\cCn$.
	\item It is unitary, with respect to the hermitian structure defined by
	\be
	\label{EqHermStrucQuantLine}
  (\psi_1, \psi_2) = \int_{\mathcal{A}} \omega^n (\psi_1, \psi_2)_{\mathscr{L}} = \int_{\mathcal{A}} \omega^n \exp(-h) \psi_1 \bar{\psi_2} \;,
  \ee
  for $\psi_1$, $\psi_2$ elements of $\mathscr{H}$.
  \item The connection form $Q$ is of type $(1,0)$ as an ${\rm End}(\mathscr{L})$-valued 1-form on $\cCn$. This fact can be explained as follows. Recall that the antiholomorphic directions on $\cCn$ parameterize the deformations of antiholomorphic functions on $\mathcal{A}$. As the quantum bundle $\cCn$ is built out of the holomorphic sections of $\mathscr{L}$, transport in the antiholomorphic directions of $\cCn$ should act trivially on $\mathscr{H}$, so the connection form has to be of type $(1,0)$.
\end{itemize}
For proofs of these three properties, see \cite{Axelrod:1989xt}. We will repeat them in the slightly more general case considered next.

\subsection{Symplectic reduction and non-trivial hermitian structures}

\label{SecGen}

We would like to generalize the construction of the projectively flat connection $\delta_{\mathscr{H}}$ to the case when the affine symplectic space $\mathcal{A}$ comes from the symplectic reduction of a (possibly infinite dimensional) affine symplectic space $\tilde{\mathcal{A}}$ through the action $\rho$ of a Lie group $\mathcal{G}$.

We denote by $\tcC$ the space of affine complex structures on $\tilde{\mathcal{A}}$. As $\tilde{\mathcal{A}}$ is an infinite dimensional space, a precise definition of $\tcC$ has to include a description of its topology.  There exists a construction of an infinite dimensional analog of the Siegel upper half-plane \cite{springerlink:10.1007/BF01208274}, but it is unclear to us if it contains the set of polarizations we are interested in, namely the polarizations on the space of $2\ell + 1$-forms obtained from Hodge star operators. In the next section, we will propose a precise definition of $\tcC$ in this case, together with a topology making it contractible. For now, we will only need the fact that $\tcC$ has a complex structure, which is automatic as it is itself a moduli space of complex structures (see the case of $\cCn$ in the previous section).

Recall that we can associate a moment map to the action of $\mathcal{G}$, namely a map $F: \tilde{\mathcal{A}} \rightarrow \mathfrak{g}^\ast$ such that $(t,\delta_{\tilde{\mathcal{A}}}F(v)) = \omega(v,\rho(t))$. In the previous equation, $\mathfrak{g} := {\rm Lie}(\mathcal{G})$, $t \in \mathfrak{g}$, $(.,.)$ is the natural pairing between $\mathfrak{g}$ and $\mathfrak{g}^\ast$, and $\delta_{\tilde{\mathcal{A}}}$ is the differential on $\tilde{\mathcal{A}}$. The symplectic reduction of $\tilde{\mathcal{A}}$ by $\mathcal{G}$ is the space of orbits of $\mathcal{G}$ on $F^{-1}(0)$. 

Let us pick $\tilde{J} \in \tcC$. This choice induces metric $g$ on $\tilde{\mathcal{A}}$: $g(v_1,v_2) = \omega(v_1, \tilde{J}v_2)$, $v_1,\, v_2 \in T\tilde{\mathcal{A}}$. We decompose the moment map on a basis dual to a basis $\{t_\alpha\}$ of $\mathfrak{g}$: $F = F_\alpha t_\alpha^\ast$. By the definition of the moment map, any vector $v \in TF^{-1}(0)$ satisfies
\be
0 = \delta_{\tilde{\mathcal{A}}} F_\alpha(v) = \omega(v, \rho(t_\alpha)) = -g(v, \tilde{J} \rho(t_\alpha)) \;,
\ee
so $TF^{-1}(0)$ is the orthogonal complement of $J\rho(\mathfrak{g})$. We can therefore identify $T\mathcal{A}$ with the complement in $TF^{-1}(0)$ of $\rho(\mathfrak{g})$ and we get the decomposition:
\be
T\tilde{\mathcal{A}} = \rho(\mathfrak{g}) \oplus J\rho(\mathfrak{g}) \oplus T\mathcal{A} = \rho(\mathfrak{g}_{\mathbbm{C}}) \oplus T\mathcal{A} \;.
\ee 
As $\tilde{J}$ is an orthogonal transformation and leaves $\rho(\mathfrak{g}_{\mathbbm{C}})$ invariant, it projects down to a complex structure on $T\mathcal{A}$. Therefore, we get a map $\pi: \tcC \rightarrow \cCn$.

We now come to an important point. The measure on $\tcA$ induces a natural measure on $\cA$ through the symplectic reduction. At a point of $\cA$, this measure is proportional to the regularized volume of the corresponding orbit of $\cG$ in $F^{-1}(0)$. The induced measure determines the Hermitian structure on the quantum line bundle through \eqref{EqHermStrucQuantLine}. In general, as a function on $\tcC$, it \emph{does not} factor through $\cCn$. As a result, the quantum bundle, its Hermitian structure and the projective connection have to be constructed on $\tcC$.

The quantum bundle $\tsH$ over $\tcC$ is defined as 
the pull-back by $\pi$ of the bundle of holomorphic sections of the prequantum bundle $\mathscr{L}$. Let us describe the non-trivial measure by a function $u: \cA \times \tcC \rightarrow \mathbbm{R}_+$. The Hermitian structure on $\tsH$ is given by 
\be
\label{EqHermStrucNT}
  (\psi_1, \psi_2)_{\tilde{J}} = \int_{\mathcal{A}} \omega^{n} (\psi_1, \psi_2)_{\mathscr{L}} u(\tilde{J}) \;,
\ee
for $\tilde{J} \in \tcC$, $\psi_1, \; \psi_2 \in \tsH_{\tilde{J}}$.

Let us now make the important assumption that $u(\tilde{J})$ is constant along $\mathcal{A}$. This is the case we will encounter in the next section in the case of the self-dual abelian gauge field. The generalization to the case when this assumption is not valid has been described to some extent in \cite{Axelrod:1989xt}. 

Recall that for a consistent geometric quantization, we have four requirement for the connection $\delta_{\tsH}$, as explained at the end of the previous section: $\delta_{\tsH}$ should preserve holomorphicity, be projectively flat, be unitary and admit a purely $(1,0)$ connection form. It turns out that these properties fully determines $\delta_{\tsH}$:
\begin{equation}
\label{EqProjConnCorr}
\delta_{\tsH} = \delta - Q \;, \qquad Q = - \frac{1}{4}(\delta J \omega^{-1})^{\underline{i} \underline{j}} \nabla_{\underline{i}} \nabla_{\underline{j}} - \delta^{(1,0)} \ln u \;,
\end{equation}
where $J = \tilde{J}|_{T\mathcal{A}}$ and by a slight abuse of notation, we are writing now $\delta$ for the differential on $\tilde{\mathcal{C}}$. Let us now check them in turn.

\subsubsection*{Holomorphicity preservation}

Write $\omega = \omega_{ij} da^i da^j$ where $a^i$ are affine coordinates on $\mathcal{A}$ and observe that the fact that $J$ is compatible with $\omega$ reads
\be
\omega(Jv_1, Jv_2) = \omega(v_1,v_2) \quad \Rightarrow \quad \omega_{ik} J^k_{\;j} = - J^k_{\;i} \omega_{kj} = \omega_{jk} J^k_{\;i} \;.
\ee 
so $\omega J$ and $\omega \delta J$ are represented by symmetric matrices. Writing $P_\pm = \frac{1}{2}(1 \mp iJ)$ for the projectors on the holomorphic and antiholomorphic tangent bundles, we get $\delta P_- = \frac{i}{2}\delta J$. We can compute
\begin{align}
[P_{-\;i}^j \nabla_j, \delta_{\tsH}] =& \, - (\delta P_-)^j_{\;i} \nabla_j + \frac{1}{4} (\delta J \omega^{-1})^{\underline{k} \underline{l}} [P_{-\;i}^j \nabla_j, \nabla_{\underline{k}} \nabla_{\underline{l}} ] \notag \\
=&\, - \frac{i}{2}(\delta J)^{\underline{j}}_{\;i} \nabla_{\underline{j}} - \frac{i}{4} (\omega \delta J \omega^{-1} - \delta J)^{\underline{j}}_i \nabla_{\underline{j}} = 0 \;,
\end{align}
where we used the fact that the curvature of $\nabla$ is $-i\omega$, as well as the symmetry of $\omega \delta J$.

\subsubsection*{Projective flatness}

As the holomorphic components of $\nabla$ commute, $Q \wedge Q = 0$. We can compute the curvature of $\delta_{\tsH}$ as follows:
\begin{align}
\label{EqCourbDH}
\delta_{\tsH}^{\;2} =& \, -\delta(Q) = \frac{1}{4} \delta (P_+ \delta J \omega^{-1})^{ij} \nabla_i \nabla_j + \delta^{(0,1)} \delta^{(1,0)} \ln u \notag \\
=& \, -\frac{i}{16} (\delta J \, \delta J \omega^{-1})^{ij} [\nabla_{i}, \nabla_{j}] + \delta^{(0,1)} \delta^{(1,0)} \ln u \notag \\
=& \, \frac{1}{16} {\rm Tr} (\delta J \delta J) + \delta^{(0,1)} \delta^{(1,0)} \ln u \;,
\end{align}
where we used
\be
(\delta J \, \delta J \omega^{-1})^{ij} = -(\delta J \, \delta J \omega^{-1})^{ji}
\ee
to extract a commutator. From \eqref{EqCourbDH}, we see that the curvature $(\delta_{\tsH})^2$ acts by scalar multiplication on the fibers of $\tsH$, so $\delta_{\tsH}$ is projectively flat.

\subsubsection*{Unitarity}

$\delta_{\tsH}$ is unitary with respect to $(.,.)$ if
\be
\label{EqCondUnitarity}
\delta (\psi_1, \psi_2) = (\delta_{\tsH} \psi_1, \psi_2) + (\psi_1, \delta_{\tsH} \psi_2) \;.
\ee
Using the expression \eqref{EqHermStrucNT} for the Hermitian structure, we get for \eqref{EqCondUnitarity} after canceling on both sides the terms involving $\delta \psi_1$ and $\delta \psi_2$:
\begin{align}
\label{EqCheckUn}
\int_{\mathcal{A}} \omega^{n} (\psi_1, \psi_2)_{\mathscr{L}} \, \delta u = &\; \int_{\mathcal{A}} \omega^{n} \left(\frac{1}{4}(\delta J \omega^{-1})^{\underline{i} \underline{j}} \nabla_{\underline{i}} \nabla_{\underline{j}} \psi_1, \psi_2 \right)_{\mathscr{L}} u \, + \rm{h.c.} \notag \\
&\; + \, \int_{\mathcal{A}} \omega^{n} (\psi_1, \psi_2)_{\mathscr{L}} \, u \, \big(\delta^{(1,0)} \ln u + \delta^{(0,1)} \ln u \big) \;,
\end{align}
h.c. denoting the hermitian conjugate of the first term of the right-hand side. As $(.,.)_{\mathscr{L}}$ is compatible with $\nabla$, we can use integration by part on the first term of the right-hand side, which vanishes because $\psi_2$ is holomorphic. The hermitian conjugate term vanishes as well. Finally, the remaining terms on the second line coincide with the left-hand side. Hence the equality is verified and $\delta_{\tsH}$ is unitarity. \vspace{.5cm}

Moreover it is obvious that the connection form $Q$ as defined in \eqref{EqProjConnCorr} is of type $(1,0)$. The connection $\delta_{\tsH}$ therefore has the four properties required for a consistent geometric quantization.

\section{The self-dual field}

\label{SecSDF}

We are now ready to construct the connection $\delta_{\tsH}$ in the case relevant to the quantization of the self-dual field. In the section \ref{SecSympRed}, we identify the ingredients of the construction of the last section in the case of the self-dual field. The explicit form of the connection is computed in section \ref{SecExplComp}. In this derivation, we disregard global issues associated to large gauge transformations and large diffeomorphisms that are irrelevant for the derivation of the local form \eqref{EqProjConnCorr} of the connection. We include these global considerations in section \ref{SecGlobIss}. We focus here on the geometric quantization problem and refer the reader to the papers \cite{Witten:1996hc, Belov:2006jd} for the physical motivation of this construction.


\subsection{Symplectic reduction on the space of $2\ell + 1$ forms}

\label{SecSympRed}

We consider a $2\ell$-form gauge field with self-dual field strength on a compact $4\ell + 2$ manifold $M$. The partition function of the self-dual field is constructed holographically in \cite{Witten:1996hc, Belov:2006jd} as the wave function of a $(2\ell + 1)$-form abelian spin Chern-Simons theory. The restriction of the Chern-Simons field on the manifold $M$ is identified with the background gauge field $A$ coupling to the self-dual field, and the wave function of the Chern-Simons theory gives the partition function of the self-dual field. We refer the reader to the two papers \cite{Witten:1996hc, Belov:2006jd} for a detailed exposition of these ideas. Practically, we will construct the wave function by geometric quantization of the space gauge fields $A$ on $M$. We will see that the abstract formalism developed in the previous section is perfectly suited to solve this problem. Note that we are considering here the case where the level $k$ of the Chern-Simons theory is equal to 1 in the conventions of \cite{Belov:2006jd}.

\subsubsection*{Gauge fields as differential cocycles}

We first have to determine the space of background gauge fields $A$. Gauge fields with possibly non-trivial topology are best described in the formalism of differential cohomology \cite{springerlink:10.1007/BFb0075216, hopkins-2005-70} (see section 2 of \cite{Freed:2006yc} for a pedagogical introduction).

Let $C^p(M,k)$ denote the space of smooth \v{C}ech cochains valued in $k$. Recall that a differential $p$-cochain is a triplet
\be
(c,a,\omega) \in C^p(M,\mathbbm{Z}) \times C^{p-1}(M,\mathbbm{R}) \times \Omega^p(M) \;.
\ee
One can define a differential
\be
d(c,a,\omega) := (\delta c, \omega - \delta a - c, d\omega) \;.
\ee
In accordance with the usual terminology, differential $p$-cocycles are elements of the kernel of $d$. We denote by $\check{C}^p(M)$ the space of differential $p$-cochains and by $\check{Z}^p(M)$ the space of differential $p$-cocycles. Gauge $(p-1)$-form fields on $M$ are elements of $\check{Z}^{p}(M)$, the component in $\Omega^p(M)$ being the field strength of the gauge field. A gauge transformation is given by the addition of the differential of an element of $\check{C}^{p-1}(M)$ with vanishing field strength. 

The space of gauge fields on our manifold $M$ is $\check{Z}^{2\ell+2}(M)$. It has an infinite number of connected components labeled by the class of $c$ in $H^{2\ell+2}(M,\mathbbm{Z})$. In each connected component, we have a subspace parameterized by the closed form $\omega$. As $\omega  - c$ is exact, the De Rahm cohomology class of $\omega$ is fixed by the class of $c$. This subspace is therefore an affine space modeled on $\Omega^{2\ell+2}_{\rm exact}(M)$, the space of exact forms of degree $2\ell+2$. For fixed $c$ and $\omega$, $a$ is fixed modulo elements in $Z^{2\ell+1}(M,\mathbbm{R}) \simeq \Omega^{2\ell+1}_{\rm closed}(M)$. As exacts forms of degree ${2\ell+2}$ are in bijection with coexact forms in degree ${2\ell+1}$, we find that each connected component in $\check{Z}^{2\ell+2}(M)$ is an infinite dimensional affine space modeled on $\Omega^{2\ell+1}(M)$. 

Let us now investigate the structure of the gauge group $\mathcal{G}$. Exact elements in $\check{C}^{2\ell+2}(M)$ with vanishing field strength are of the form $(\delta c, - \delta a - c, 0)$. $- \delta a - c$ is simply a real cocycle with integral periods, or equivalently an element of $\Omega_{\mathbb{Z}}^{2\ell + 1}(M)$, the set of $(2\ell+1)$-forms with integral periods. Therefore $\mathcal{G} \simeq \Omega_{\mathbb{Z}}^{2\ell + 1}(M)$. The connected component of the identity is $\mathcal{G}_0 \simeq \Omega_{\rm exact}^{2\ell + 1}(M) \simeq \Omega_{\rm coexact}^{2\ell}(M)$. It is associated with small gauge transformations. For $\xi \in \Omega_{\mathbb{Z}}^{2\ell + 1}(M)$ and $A \in \check{Z}^{2\ell+2}(M)$, the action of the gauge group is simply the affine transformation 
\be
\label{EqActionGA}
A \rightarrow A - \xi \;.
\ee

We make a choice for the integral cohomology class $c$ of the field strength of the background field, which selects a single component $\tcA$ in $\check{Z}^{2\ell+2}(M)$. Physically, this amounts to specifying a flux background on $M$. $\tcA$ is the affine space we would like to quantize. The symplectic form on $\tilde{\mathcal{A}}$ reads
\be
\label{EqSympForm}
\omega(\phi_1,\phi_2) = \pi \int_{M} \phi_1 \wedge \phi_2 \;,
\ee
where $\phi_1, \phi_2 \in T_A \tilde{\mathcal{A}} \simeq \Omega^{2\ell + 1}(M)$ are tangent vectors at $A \in \tilde{\mathcal{A}}$. Note that this form is compatible with the affine structure on $\tilde{\mathcal{A}}$.

\subsubsection*{Moment map}

The action \eqref{EqActionGA} preserves the symplectic form \eqref{EqSympForm}. We will temporarily ignore large gauge transformation and consider only the action of $\mathcal{G}_0$ on $\tcA$.

Let us find the moment map for the action of $\mathcal{G}_0$. 
We set $\mathfrak{g} = Lie(\mathcal{G}_0) \simeq \Omega^{2\ell}_{\rm coexact}(M)$. The tangent vector corresponding to the infinitesimal action of $\epsilon \in \mathfrak{g}$ is $-d\epsilon$. Recall that the moment map is a function $F : \tilde{\mathcal{A}} \rightarrow \mathfrak{g}^\ast$ satisfying 
\be
\label{EqMomMapSD}
\omega(-d \epsilon, \phi) = \delta_{\tilde{\mathcal{A}}}(F(A),\epsilon)(\phi) \;.
\ee
As $\mathfrak{g} \simeq \Omega^{2\ell}(M)$, $\mathfrak{g}^\ast \simeq \Omega^{2\ell+2}(M)$ by Poincar\'e duality. We can therefore rewrite \eqref{EqMomMapSD} more explicitly:
\be
\label{EqMomMapSDExpl}
-\pi \int_M d\epsilon \wedge \phi = \delta_{\tilde{\mathcal{A}}}\left( \int_M \epsilon \wedge F(A) \right)(\phi) =  \phi \left(\int_M \epsilon \wedge F(A)\right) \;,
\ee
where in the second equality, we used the definition of the differential $\delta_{\tilde{\mathcal{A}}}$. Let us write $A = (c,a,\omega)$, and decompose $\omega = \omega_0 + \omega_A$, where $\omega_0$ is the harmonic form satisfying $\omega_0 - c = 0$ and $\omega_A$ has a trivial image in cohomology. Then the infinitesimal variation of $A$ generated by the vector $\phi$ changes $\omega_A$ to $\omega_A + d\phi$. Therefore a simple integration by part shows that $F(A) = \omega_A$ solves \eqref{EqMomMapSDExpl}. We recovered the well-known fact that the moment map associated to gauge transformations is given by the field strength of the gauge field, generalized here to the case of gauge fields with non-trivial topology.

\subsubsection*{Symplectic reduction}

The equations of motion of Chern-Simons theory impose $F(A) = \delta_{\rm sources}$, where $\delta_{\rm sources}$ is a closed $(2\ell + 2)$-form accounting for the possible sources for the background gauge field on $M$. The equations of motion are invariant under the action of the gauge group $\mathcal{G}_0$, so the system can be reduced to $\mathcal{A} = F^{-1}(\delta_{\rm sources})/\mathcal{G}_0$. The preimage of $F^{-1}(\delta_{\rm sources})$ is a torsor on $\Omega^{2\ell+1}_{\rm closed}(M)$. As $\mathcal{G}_0 \simeq \Omega^{2\ell+1}_{\rm exact}(M)$, it is clear that $\mathcal{A}$ is a torsor on $\mathcal{H}^{2\ell + 1}(M)$, the space of harmonic $(2\ell+1)$-forms on $M$. Therefore it is a finite dimensional affine space.


\subsubsection*{Complex structures on $\mathcal{A}$}

The choice of a metric $g$ on $M$ provides a Hodge star operator $\ast$ acting on $\Omega^{2\ell + 1}(M) = T \tilde{\mathcal{A}}$. In dimension $4\ell + 2$, the Hodge star squares to $-\mathbbm{1}$ on $\Omega^{2\ell + 1}(M)$, thereby providing a complex structure on $\tilde{\mathcal{A}}$.

We must now make the definition of the moduli of complex structures $\tcC$ more precise. To this end, we can take advantage of the fact that the Hodge star operator commutes with the Laplacian, and that the eigenvalues of latter defines a natural filtration on $\tilde{\mathcal{A}}$. For each $\lambda \in \mathbbm{R}_+$ of the Laplacian, define $\mathcal{A}_{\lambda}$ to be the finite dimensional space of eigenforms with eigenvalue less than $\lambda$. Let $\cC_\lambda$ be the Siegel upper-half plane of polarizations of $T\mathcal{A}_{\lambda}$ and let us choose a reference polarization on $\tilde{\mathcal{A}}$ generated by some metric on $M$. Given a polarization on $\mathcal{A}_{\lambda}$, we can complete it to a polarization of $\mathcal{A}_{\lambda'}$ for any $\lambda < \lambda'$, using the reference polarization on $\mathcal{A}_{\lambda'}/\mathcal{A}_{\lambda}$. Therefore we have inclusions $\cC_\lambda \subset \cC_{\lambda'}$ for any $\lambda < \lambda'$. Now define $\tcC$ as the direct limit of the family $\{\cC_\lambda\}_{\lambda \in \mathbbm{R}_+}$. As a direct limit of contractible spaces, $\tcC$ is contractible as well. While it is not completely clear whether $\tcC$ is independent of the choice of reference polarization or not, it is clear that it contains all the polarizations that can be obtained from metrics on $M$ through the associated Hodge star operator. This is sufficient for our purpose. 

The Hodge star operator provides a map from the space $\mathcal{M}$ of all Riemannian metrics on $M$ into the infinite period domain $\tcC$. We also saw in the previous section that there exists a projection map $\pi : \tcC \rightarrow \cCn$ onto the finite dimensional period domain parameterizing the complex structures on the symplectically reduced phase space $\mathcal{A}$. In our case, the map $\pi$ simply corresponds to the restriction of the Hodge star operator to the space of harmonic forms $\mathcal{H}^{2\ell + 1}(M)$.

\subsubsection*{Measure on the symplectic reduction}

We endow $\tilde{\mathcal{A}}$ with a constant measure. This measure pushes down to a measure on $\mathcal{A}$, after restriction to $F^{-1}(\delta_{\rm sources})$ and integration along the orbits of $\mathcal{G}_0$. The induced measure was computed in the appendix C of \cite{Belov:2006jd} (see also section 3 of \cite{Monnier2011}) and is given by
\be
\mathcal{D}a^H u(g)
\ee
with
\be
\label{EqMesSR}
u^2(g) = \prod_{p = 0}^{2\ell} \left ( \left ( V_p^{-2} {\rm det}'(d^\dagger d|_{\Omega^p(M) \cap {\rm Im}(d^\dagger)}) \right )^{(-1)^p} \right )\;,
\ee
where $\mathcal{D}a^H$ is the measure coming from the natural $L^2$ norm on harmonic $2\ell + 1$-forms, $V_p$ is the volume of the torus of harmonic $p$-forms and ${\rm det}'$ denotes the zeta-regularized determinant. Because of these determinants, $u$ is a function on $\tcC$ that does not push down to a function on $\cCn$.

Let us briefly recall how the zeta regularized determinant is defined (see for instance section 9.6 of \cite{MR1215720} or chapter 5 of \cite{Rosenberg1997}). 
The zeta function associated to ${\rm det}'(d^\dagger d)_p$ is defined by:
\be
\label{EqDefZeta}
\zeta_p(s) = \sum_{\lambda \in {\rm Spec}_p} \lambda^{-s} \;,
\ee
for ${\rm Re}(s) > 1$, where we wrote ${\rm Spec}_p$ for the spectrum of $d^\dagger d|_{\Omega^p(M) \cap {\rm Im}(d^\dagger)}$. 
$\zeta$ admits a meromorphic extension to the whole complex plane and is holomorphic at $s = 0$. The regularized determinant is
\be
{\rm det}'(d^\dagger d|_{\Omega^p(M) \cap {\rm Im}(d^\dagger)}) = e^{-\zeta_p'(0)} \;,
\ee
where $\zeta_p'$ is the derivative of $\zeta_p$. From \eqref{EqDefZeta}, it is clear that $\zeta(s)$ is real for $s$ real and larger than $1$. As a result, $\zeta_p(s)$ is real on the whole real axis (except at possible singularities), so $\zeta_p'(0)$ is real as well. $u^2(g)$ is therefore valued in $\mathbbm{R}_+$, $u(g)$ can be taken as the positive square root and it defines a constant measure on $\mathcal{A}$.\\

Now we can apply the treatment of section \ref{SecGen}, with the same notation. We get a connection \eqref{EqProjConnCorr} on the quantum bundle that is projectively flat and unitary.

\subsection{Explicit computation of the projective connection}

\label{SecExplComp}

In this section, we pick a coordinate system on $\mathcal{A}$ and a trivialization of $\mathscr{L}$ to compute explicitly the connection $\delta_{\tsH}$.

\subsubsection*{The coordinate system}

We choose a base point in $\mathcal{A}$ and get an isomorphism of $\mathcal{A}$ with the space of harmonic forms $\mathcal{H}^{2\ell + 1}$. We denote the lattice $\mathcal{H}^{2\ell + 1}_\mathbb{Z}(M)$ of integral harmonic forms by $\Lambda$. We choose a Lagrangian decomposition $\mathcal{H}^{2\ell + 1} = V_1 \oplus V_2$, which induces a decomposition $\Lambda = \Lambda_1 \oplus \Lambda_2$. We also pick a basis $\{\alpha_i\} \in \Lambda_1$, $\{\beta^i\} \in \Lambda_2$ satisfying
\be
\omega(\alpha_i, \alpha_j) = \omega(\beta^i,\beta^j) = 0 \qquad \omega(\alpha_i,\beta^j) = 2 \pi \delta_i^j \;.
\ee
The complex structure induced by the metric through the Hodge star operator is characterized by a period matrix $\tau$. By definition, the holomorphic (constant) vectors fields are linear combinations of the vectors
\be
\zeta_{\underline{i}} = \alpha_i + \bar{\tau}_{ij} \beta^j \;.
\ee
It may seem like an awkward choice that the holomorphic vector fields should depend on $\tau$ antiholomorphically. However, with this choice the holomorphic coordinates \eqref{EqDefz} depend holomorphically on $\tau$. We introduce coordinates $a^i$, $b_i$ such that 
\be
\alpha_i = \frac{\partial}{\partial a^i} \qquad \beta^i = \frac{\partial}{\partial b_i} \;.
\ee
Let us define the metric $h_{ij} = -i(\tau - \bar{\tau})_{ij}$. We then have $\zeta_i = h_{ij} \frac{\partial}{\partial z_j}$, with 
\be
\label{EqDefz}
z_j = -i(\tau_{jk} a^k - b_j) \qquad \bar{z}_j = i(\bar{\tau}_{jk} a^k - b_j) \;.
\ee
We also define $h^{ij} = i\big((\tau-\bar{\tau})^{-1}\big)^{ij}$. When indices are omitted, this matrix is denoted by $h^{-1}$. Then
\be
a^i = h^{ik}(z_k + \bar{z}_k) \;, \qquad b_i = \bar{\tau}_{ij}h^{jk}z_k + \tau_{ij}h^{jk} \bar{z}_k \;.
\ee
$\omega$ can is expressed as follows:
\be
\omega = 2\pi da^i \wedge db_i = 2\pi i h^{ij} dz_i \wedge d\bar{z}_j \;.
\ee
The total derivatives with respect to the matrix elements of $\tau$ are given by:
$$
\frac{d}{d\tau_{ij}} = \frac{\partial}{\partial \tau_{ij}} - i h^{ik}(z_k + \bar{z}_k) \frac{\partial}{\partial z_j} \;,
$$
\be
\label{EqTotDifTau}
\frac{d}{d\bar{\tau}_{ij}} = \frac{\partial}{\partial \bar{\tau}_{ij}} + i h^{ik}(z_k + \bar{z}_k) \frac{\partial}{\partial \bar{z}_j} \;.
\ee
where the partial derivative $\frac{\partial}{\partial \tau_{ij}}$ is taken with $z_i$ and $\bar{z}_i$ held constant. $\tau$ is a symmetric matrix: $\tau_{ij} = \tau_{ji}$. To avoid clumsy summation signs, we use the convention
\be
\delta \tau_{ij} A^{i \langle j} = \sum_{i \leq j} \delta \tau_{ij} A^{ij} \;.
\ee
The space of affine complex structure $\tcC$ on $\Omega_{\mathbbm{C}}^{2\ell + 1}(M)$ is parameterized by $\tau_{ij}$, $\bar{\tau}_{ij}$ as well as an infinite set of extra holomorphic and antiholomorphic coordinates that we write $\tau_{\alpha}$ and $\bar{\tau}_{\alpha}$, respectively.

\subsubsection*{Trivialization}

The explicit form of $\delta_{\tsH}$ can take many equivalent forms, each related to a choice of trivialization of the prequantum bundle $\mathscr{L}$. With the most natural choice, the data on the prequantum bundle reads:
\begin{align}
\label{EqSimChTriv}
&\nabla^{\underline{i}} = \frac{\partial}{\partial z_i} - \pi h^{ij}\bar{z}_j \;, \notag \\
&\nabla^{\overline{i}} = \frac{\partial}{\partial \bar{z}_i} + \pi h^{ij} z_j \;, \notag \\
&(\psi_1, \psi_2)_{\mathscr{L}} = \psi_1 \bar{\psi}_2 \;, \\
&\delta  = \delta \tau_{ij} \frac{d}{d \tau_{i \langle j}} + \delta \bar{\tau}_{ij} \frac{d}{d \bar{\tau}_{i \langle j}} + \delta \tau_{\alpha} \frac{d}{d \tau_{\alpha}} + \delta \bar{\tau}_{\alpha} \frac{d}{d \bar{\tau}_{\alpha}} \;, \notag
\end{align}
As required, $\nabla$ has curvature $-i\omega$.

It is useful to find a trivialization in which the holomorphic sections of $\mathscr{L}$ are independent of the coordinates $\bar{z}_j$. This amounts to requiring that $\nabla^{\overline{i}} = \frac{\partial}{\partial \bar{z}_i}$. This is realized by the change of trivialization $\psi \rightarrow s \psi$, for $\psi$ a section of $\mathscr{L}$ and
\be
s(\tau, z, \bar{z}) = \exp \left( \pi z_i h^{ij} (z_j + \bar{z}_j)\right) \;.
\ee
Performing the corresponding gauge transformation $(\nabla,\, \delta) \rightarrow (s \nabla s^{-1},\, s \delta s^{-1})$, we get:
$$
\nabla^{\overline{i}} = \frac{\partial}{\partial \bar{z}_i} \;, \qquad 
\nabla^{\underline{i}} = \frac{\partial}{\partial z_i} - 2\pi a^i \;,
$$
\be
\label{EqDeltaTriv2}
\delta  = \delta \tau_{ij} \left(  \frac{d}{d \tau_{i \langle j}} + \pi i a^i  a^j \right) + \delta \bar{\tau}_{ij} \frac{d}{ d \bar{\tau}_{i \langle j}} + \delta \tau_{\alpha} \frac{d}{d \tau_{\alpha}} + \delta \bar{\tau}_{\alpha} \frac{d}{d \bar{\tau}_{\alpha}} \;.
\ee

\subsubsection*{The one-form $\delta J$ on $\mathcal{T}$}

We have $J \bar{\zeta}_j = -i \bar{\zeta}_j$, $\bar{\zeta}_j = \alpha_j + \tau_{jk} \beta^k$. Therefore
\be
\delta J \bar{\zeta}_j = - J \delta (\bar{\zeta}_j) - i \delta (\bar{\zeta}_j) = -2i P_+ \delta \tau_{jk} \beta^k =  2 \delta \tau_{jk} h^{kl} \zeta_l \;,
\ee
where we used 
\be
\beta^k = i h^{kl} (\zeta_l - \bar{\zeta}_l) \;.
\ee
We deduce
\be
\delta J_{\; \overline{j}}^{\underline{i}} = 2 h^{ik} \delta \tau_{kj} \;. 
\ee

\subsubsection*{The connection form $Q$}

Recall that
\be
Q = - \frac{1}{4}(\delta J \omega^{-1})^{\underline{i} \underline{j}} \nabla_{\underline{i}} \nabla_{\underline{j}} - \delta^{(1,0)} \ln u \;.
\ee
To compute $\omega^{-1}$, we have to express $\omega$ in the basis $\{dz^i, d\bar{z}^i\}$ dual to $\{\zeta_i, \bar{\zeta}_i\}$: $\omega = 2\pi i h_{ij}dz^i \wedge d\bar{z}^j$. We have therefore:
\be
(\delta J \omega^{-1})^{\underline{i} \underline{j}} = -\frac{i}{\pi} (h^{-1} \delta \tau h^{-1})^{ij} \;,
\ee
so 
\begin{align}
\label{EqQ}
Q \,+ \delta^{(1,0)} \ln u = & \; \frac{i}{4\pi} \delta \tau_{ij} \nabla^{\underline{i}}\nabla^{\underline{j}}  \notag \\
=& \; \delta \tau_{ij} \left ( \frac{i}{4\pi} \frac{\partial}{\partial z_i} \frac{\partial}{\partial z_j} - i a^i \frac{\partial}{\partial z^j} - \frac{i}{2} h^{ij} + \pi i a^i a^j  \right ) \;.
\end{align}

\subsubsection*{Formula for $\delta_{\tsH}$}

Replacing \eqref{EqTotDifTau}, \eqref{EqDeltaTriv2} and \eqref{EqQ} in the definition \eqref{EqProjConnCorr}, we get an explicit form for the connection on the quantum bundle:
\begin{align}
\label{EqConnExpl}
\delta_{\tsH} =& \;\; \delta_{\tsH}^{(1,0)} + \delta_{\tsH}^{(0,1)} \notag \\
\delta_{\tsH}^{(0,1)} =& \;\; \delta^{(0,1)} = \delta \bar{\tau}_{ij} \frac{d}{d \bar{\tau}_{i \langle j}} + \delta \bar{\tau}_{\alpha} \frac{d}{d \bar{\tau}_{\alpha}} \notag \\
\delta_{\tsH}^{(1,0)} =& \;\; \delta^{(1,0)} - Q  \\
= & \;\; \delta \tau_{ij} \left ( \frac{\partial}{\partial \tau_{i \langle j}} - ia^i \frac{\partial}{\partial z^j} + \pi i a^i a^j \right ) - \;  Q + \delta \tau_\alpha \frac{d}{ d \tau_\alpha} \notag \\
= & \;\; \delta \tau_{ij} \left ( \frac{\partial}{\partial \tau_{i \langle j}} - \frac{i}{4\pi} \frac{\partial}{\partial z_i} \frac{\partial}{\partial z_j} + \frac{i}{2} h^{ij}  \right ) + \delta \tau_\alpha \frac{\partial}{\partial \tau_\alpha} + \delta^{(1,0)} \ln u \;. \notag
\end{align}
It might be puzzling to the reader that a connection should take the form of a second order differential operator. One should remember that the quantum bundle has as fiber the space of holomorphic sections of the prequantum bundle. The $z$-dependent second order differential operator is a linear operator acting on holomorphic sections, and once a basis for the latter is chosen, one gets a more familiar matrix-valued connection, as we will see explicitly later. 

Let us note that $\delta_{\tsH}$ differs by a $1$-form from the pull-back by $\pi$ of the following connection
\be
\label{EqConnDown}
\delta^{(1,0)}_{\sH} = \delta \tau_{ij} \left ( \frac{\partial}{\partial \tau_{i \langle j}} - \frac{i}{4\pi} \frac{\partial}{\partial z_i} \frac{\partial}{\partial z_j} + \frac{i}{2} h^{ij}  \right ) \;, \quad \delta^{(0,1)}_{\sH} = \delta \bar{\tau}_{ij} \frac{d}{d \bar{\tau}_{i \langle j}}
\ee
on a bundle $\sH$ on $\cCn$. $\delta_{\tsH}$ is therefore gauge equivalent to the pull-back of $\delta_{\sH}$ to $\tcC$. However, the one-form $\delta^{(1,0)} \ln u$ is not the pull-back of a one-form on $\cCn$, therefore the partition function itself does not pull-back from a section of $\sH$. Because of the obvious convenience of working with bundles over a finite dimensional space, we will use $\delta_{\sH}$ to study topological questions about the partition function.

\subsection{Global issues}

\label{SecGlobIss}

So far, our derivation of the connection on the quantum bundle has been purely local on $\mathcal{A}$ and on $\mathcal{M}$, the manifold of Riemannian metrics on $M$. To simplify the discussion, we ignored two global issues that were irrelevant for the local derivation of $\delta_{\tsH}$. We now would like to discuss them.

\subsubsection*{Large gauge transformations}

The first issue is that instead of the affine space $\mathcal{A}$, we should really quantize the torus $\mathcal{J} = \mathcal{A}/\Lambda$. Physically, it can be interpreted as the fact that the background field is defined up to large gauge transformations. Just as in the affine case, the Hodge star operator provides a complex structure on $\mathcal{J}$. The intersection form endows $\mathcal{J}$ with the structure of a principally polarized abelian variety. This abelian variety is known as the Lazzeri intermediate Jacobian \cite{MR1785408, MR1713785}.\footnote{\label{FNWeilLazzeri}Note that in the case when $M$ is a simply connected Calabi-Yau threefold, the Lazzeri and Weil intermediate Jacobian coincide, and the latter designation is preferred in the physics literature. It should be however emphasized that the Weil Jacobian is an abelian variety only in the case when there is no non-primitive cohomology in degree $3$ and that its complex structure requires the existence of a complex structure on $M$. In contrast, the Lazzeri Jacobian can be defined for any real oriented $4\ell + 2$ dimensional manifold and is always an abelian variety.}

The symmetric holomorphic line bundles on $\mathcal{J}$ with curvature $\omega$ are classified by a characteristic $\eta \in (\frac{1}{2}\mathbbm{Z}/\mathbbm{Z})^{2n}$ \cite{MR2062673}. We will call them $\mathscr{L}^\eta$. It is worth mentioning that the connection \eqref{EqSimChTriv} can be seen as the pull-back to $\mathcal{A}$ of a connection on $\mathscr{L}^\eta$ for any $\eta$. Indeed, a connection on the trivial line bundle $\mathscr{L}$ on $\mathcal{A}$ does not determine uniquely a connection on a bundle over $\mathcal{J}$. \footnote{This is most easily visualized in the case of flat bundles. Given any flat bundle on a manifold, one can always choose a trivialization of its pull-back on the universal cover such that the connexion form vanishes. Any global information about the bundle is lost when pulling-back to the universal cover.}

Each of the bundles $\mathscr{L}^\eta$ admits up to scalar multiples a unique holomorphic section, given by the level one theta function with the corresponding characteristic. The quantum bundle $\mathscr{H}^\eta$ is therefore a line bundle that a priori depends on $\eta$. The theta function governs the dependence of the partition function on the background $2\ell + 1$-form field. In the trivialization \eqref{EqDeltaTriv2}, they take the form of classical theta functions:
\be
\label{EqCLassThetF}
\theta^\eta(z, \tau) = \sum_{r \in \Lambda_1 - \eta_1} \exp \left ( \pi i  r^i \tau_{ij} r^j  - 2\pi i (z_k - \eta^2_k) r^k \right ) \;,
\ee
where $\eta = (\eta_1, \eta_2)$. In our setup, the characteristic is not determined by physics: it is a free parameter of the self-dual field theory. By seeing the theta function as an instanton sum, one can interpret the characteristic as a discrete theta angle \cite{Witten:1996hc}. To avoid confusion with theta functions, we will always refer to the theta angle as the ``characteristic''. A choice of characteristic is also equivalent to a choice of quadratic refinement of the intersection form (QRIF) on the middle dimensional cohomology of $M$ \cite{Belov:2006jd}.

The classical theta functions satisfy the famous heat equation:
\be
\delta \tau_{ij} \left ( \frac{\partial}{\partial \tau_{i \langle j}} - \frac{i}{4\pi} \frac{\partial}{\partial z_i} \frac{\partial}{\partial z_j} \right )\theta^\eta(z, \tau) = 0 \;,
\ee
and we can use them to trivialize the quantum bundle $\mathscr{H}^\eta$ on $\cCn$. We can write $\psi = p(\tau) \theta^\eta(z, \tau)$ for a section $\psi$ of $\mathscr{H}^\eta$, and $p$ a holomorphic function. We get in this trivialization the following simple form for the connection on $\mathscr{H}^\eta$:
\be
\label{EqConnTrivTheta}
\delta^{(1,0)}_{\sH} p = \delta \tau_{ij} \left ( \frac{\partial}{\partial \tau_{i \langle j}}  + \frac{i}{2} h^{ij}  \right )p \;, \quad \delta^{(0,1)}_{\sH}p = \delta \bar{\tau}_{ij} \frac{\partial p}{\partial \bar{\tau}_{i \langle j}} \;.
\ee
We can trivialize $\tsH$ in the same way, and the connection $\delta_{\tsH}$ then reads
\be
\label{EqConnTrivTheta_tsH}
\delta^{(1,0)}_{\tsH} p = \left( \delta^{(1,0)} + \frac{i}{2} h^{ij} \delta \tau_{ij} + \delta^{(1,0)} \ln u  \right )p \;, \quad \delta^{(0,1)}_{\sH}p = \delta^{(0,1)}p \;.
\ee

\subsubsection*{Large diffeomorphisms}

The second global issue concerns the space of Riemannian metrics $\mathcal{M}$. We expect a quantum field theory to be invariant under coordinate changes on the manifold $M$ on which it is defined. As a result, the partition function should be defined on the space of metrics $\mathcal{M}$ quotiented by the group $\mathcal{D}$ of diffeomorphisms of $M$. The self-dual field theory is more subtle, because the characteristic is not invariant under all diffeomorphisms. The best way to deal with this lack of invariance depends on the physical model in which the self-dual field theory is embedded (see \cite{Alexandrov:2010ca} for an example). As we are considering the self-dual theory for itself, this lack of invariance simply means that the partition function can be defined only on the quotient of the space of metric by the group of diffeomorphisms preserving the characteristic. Because of the existence of a gravitational anomaly, the partition function is the section of a line bundle on this quotient. Describing this line bundle is an important problem.

An important clue for the description of the anomaly line bundle comes from the fact that the action of diffeomorphisms on the space of metrics descends, through its action on the Hodge star operator, to an action on $\tcC$ and $\cCn$. The latter factorizes through the familiar action of $Sp(2n, \mathbbm{Z})$ on the Siegel upper half-plane. The subgroup of $Sp(2n, \mathbbm{Z})$ preserving all the characteristics is the level 2 congruence subgroup $\Gamma^{(2)}_{2n}$, defined as the kernel of the reduction modulo two $Sp(2n, \mathbbm{Z}) \rightarrow Sp(2n, \mathbbm{Z}_2)$. The quotient $\cTn$ of $\cCn$ by $\Gamma^{(2)}_{2n}$ is a modular variety whose Picard group is known \cite{2009arXiv0908.0555P, 2009MPCPS.147..369S}. We believe that this knowledge will be very useful for a precise description of the anomaly bundle of the self-dual field (see \cite{Monnier2011}). We will see that the connection we derived from geometric quantization allows to determine the real Chern class of this bundle. Unfortunately, there is no way to deduce the integral Chern class from geometric quantization, because the connection is defined on the universal covering $\cCn$ (see also the discussion about the prequantum line bundle in the previous section). There should be a way of deriving this information as well as the global gravitational anomaly from the evaluation on mapping tori of the spin Chern-Simons action constructed by Hopkins and Singer \cite{hopkins-2005-70}, but we leave this for future work.

As we cannot get any global information from geometric quantization, we will continue to consider the quantum bundles $\mathscr{H}^\eta$ as bundles with connection on $\cCn$, instead as bundles on $\cTn$. It is clear that, as bundles with connection on $\cCn$, they are all isomorphic to each other, but as bundles on $\cTn$ they might differ by a torsion class. We will continue to write $\mathscr{H}^\eta$ to remind the reader about this fact.

\section{Local gravitational anomaly}

\label{SecAnom}

In this section, we study the local anomaly of the self-dual field theory. The latter is related to the curvature of the connection $\delta_{\tsH^\eta}$ derived in the previous section. Topologically, we identify the inverse of the quantum bundle as a square root of the holomorphic cotangent bundle of $\cA$. However, we would like first to clarify the relation of $\tsH^\eta$ to the anomaly bundle of the self-dual field theory, that is the bundle of which the partition function of the self-dual field theory is a section of. 

The Hermitian structure of $\tsH^\eta$ can be encoded in a real function $K$ over $\tcC$, such that
\be
(\psi_1,\psi_2)_{\tsH^\eta} = \exp(-K) \psi_1 \bar{\psi_2} \;.
\ee
The function $K$, sometimes called the Kähler potential, can be extracted from the connection on $\tsH^\eta$. Indeed, in order to be compatible with the Hermitian structure, it has to take the form $\nabla_{\tsH^\eta} = \delta - \delta^{(1,0)}K$. From \eqref{EqConnTrivTheta_tsH}, using the fact that $\frac{i}{2} h^{ij} \delta \tau_{ij} = -\delta^{(1,0)} \ln \det h$, we see that $K$ is given by 
\be
K = \ln \det h - \ln u \;.
\ee
Given a polarization, the state of the $4\ell+3$-dimensional Chern-Simons theory associated with the $4\ell+2$ manifold $M$ is an element of the corresponding fiber of $\tsH^\eta$. The set of states for different choices of polarizations forms a section of $\tsH^\eta$. As it represents a state, it is natural to require the norm of this section to be constant over $\tsH^\eta$, for instance equal to 1. The normalized state of the Chern-Simons theory then reads
\be
\label{EqStatCS}
\Psi_{\rm CS} = \exp(K/2) \theta(z,\tau) \;.
\ee

Now we would like to interpret $\Psi_{\rm CS}$ as the partition function of the self-dual field theory on $M$. We expect the latter to be a holomorphic object on $\tcC$, because it depends on the metric only through the action of the Hodge star operator on self-dual forms (and not on its action on anti self-dual forms). But $\Psi_{\rm CS}$ is clearly not a holomorphic section of $\tsH^\eta$, because of the factor $\exp(K/2)$. The only way to see $\Psi_{\rm CS}$ as a holomorphic object on $\tcC$ is to see it as a section of the bundle $(\tsH^\eta)^{-1}$. Indeed, there exists a trivialization of $(\tsH^\eta)^{-1}$ in which holomorphic sections take the form of holomorphic functions multiplied by $\exp(K/2)$. In the canonical trivialization, the Hermitian norm of $(\tsH^\eta)^{-1}$ is given by
\be
(\psi_1,\psi_2)_{(\tsH^\eta)^{-1}} = \exp(K) \psi_1 \bar{\psi_2} \;.
\ee
But in the (non-holomorphic) trivialization of $(\tsH^\eta)^{-1}$ in which the Kähler potential vanishes, holomorphic sections take the form \eqref{EqStatCS}. If $\Psi_{\rm CS}$ is to be seen as a holomorphic object, it therefore has to be seen as a section of $(\tsH^\eta)^{-1}$. As a result, we learn that the anomaly bundle of the self-dual field theory is the \emph{inverse} of the quantum bundle. This subtle point, that we missed in the first version of this paper, is nicely confirmed by the path integral construction of the partition function of the self-dual field theory \cite{Monnier2011}. \\

Let us compute the curvature of $\delta_{\tsH^\eta}$. From \eqref{EqConnTrivTheta_tsH}, the curvature reads:
\be
\label{EqCurvPCb}
R(\delta_{\tsH^\eta}) = (\delta_{\tsH^\eta})^2 = -\frac{1}{2} h^{ij} h^{kl}  \delta\tau_{il} \delta\bar{\tau}_{jk} +  \delta^{(0,1)} \delta^{(1,0)} \ln u \;.
\ee
Similarly, the curvature of the connection \eqref{EqConnTrivTheta} on $\sH^\eta$ is given by
\be
\label{EqCurvPC}
R(\delta_{\sH^\eta}) = -\frac{1}{2} h^{ij} h^{kl}  \delta\tau_{il} \delta\bar{\tau}_{jk} \;.
\ee
As $R(\delta_{\tsH^\eta})$ does not vanish, the self-dual field displays a local gravitational anomaly, as was discovered a long time ago by Alvarez-Gaum\'e and Witten \cite{AlvarezGaume:1983ig}. We will make contact with their result later in the paper, when we will have identified the relevant Dirac operator. Note that the curvature is independent of $\eta$, which means that the bundles $\sH^\eta$, considered as bundles over $\cTn$, differ at most by a torsion class.

As a bundle on $\cCn$, $(\sH^\eta)^{-1}$ coincides with the square root of the determinant bundle $\mathscr{K}$ of the holomorphic cotangent space of $\mathcal{A}$. \footnote{This square root is well-known as the ``bundle of half-forms'' in the literature on geometric quantization (see for instance chapter 10 of \cite{MR1183739}).} Let us check that the curvature of $\mathscr{K}$ is given by minus twice the curvature of $\sH^\eta$. Note that $\mathscr{K}$ has an obvious global non-vanishing holomorphic section $s$ over $\cCn$, given by 
\be
\label{EqSecK}
s = d z_1 \wedge ... \wedge d z_{n} \;.
\ee
Using the explicit expression \eqref{EqDefz} of the coordinates $z_i$, we can compute the norm of $s$:
\be
|s|^2 \, v := (-i)^{n} s \wedge \bar{s} = (-i)^{n} \det (\tau - \bar{\tau}) \, v = \det(h) \, v\;,
\ee
where $v$ is the volume form $da^1 \wedge ... \wedge da^{n} \wedge db_1 \wedge ... \wedge db_{n}$. We can use $s$ to trivialize $\mathscr{K}$, and in this trivialization, the hermitian structure reads
\be
(\psi_1, \psi_2)_{\mathscr{K}} = \det(h) \; p_1 p_2 \;,
\ee
where $\psi_1 = p_1 s$, $\psi_2 = p_2 s$. The holomorphic connection compatible with this hermitian structure is $\delta_{\mathscr{K}} := \delta + \delta^{(1,0)} \ln \det(h)$, whose curvature is obviously 
\be
R(\delta_{\mathscr{K}}) = \delta^{(0,1)} \delta^{(1,0)} \ln \det(h) = h^{ij} h^{kl}  \delta\tau_{il} \delta\bar{\tau}_{jk} \;.
\ee
Therefore modulo torsion, $(\sH^\eta)^{-1}$ is a square root of $\mathscr{K}$.

\section{Metric dependence of the partition function}

\label{SecMetricDep}

This section is devoted to the study of the norm of the partition function of the self-dual field, as a function on the space of metrics. We first show that thanks to the unitarity of $\delta_{\tsH}$, only the phase of the partition function is ill-defined. 
We then demonstrate that the non-trivial metric dependence of the norm is given by the square root of the Cheeger half-torsion of $M$. 

\subsection{Normalization factor}

As was mentioned in section \ref{SecGlobIss}, the metric dependence of the partition function has to be described on the infinite dimensional space $\tcC$, rather than on $\cCn$. Recall the expression \eqref{EqConnTrivTheta_tsH} for the connection $\nabla_{\tsH^\eta}$, obtained by expressing sections of $\tsH^\eta$ as $\psi = p \, \theta^\eta$:
\be
\label{EqConnTrivTheta2}
\delta^{(1,0)}_{\tsH} p = \left(\delta^{(1,0)} + \frac{i}{2} h^{ij} \delta \tau_{ij} + \delta^{(1,0)} \ln u \right) p \;, \quad \delta^{(0,1)}_{\tsH} p = \delta^{(0,1)}p\;.
\ee
We express the Chern-Simons state $\Psi_{\rm CS}$ in the same way: $\Psi_{\rm CS} = p_{\rm CS} \, \theta^\eta$. The parallel transport of the Chern-Simons state along $\tcC$ is described by the set of partial differential equations $\delta_{\tsH} p_{\rm CS} = 0$. Of course, this set of PDE's is not integrable, because of the local anomaly described in the previous section. However, the norm of $p_{\rm CS}$ satisfies:
\be
\delta \ln |p_{\rm CS}| = \delta^{(1,0)}  {\rm Re}\left( \frac{1}{2} \ln \det(h) - \ln u \right) = \frac{1}{2}\delta \left( \frac{1}{2} \ln \det(h) - \ln u \right) \;,
\ee
which has the obvious solution:
\be
|p_{\rm CS}| = \det(h)^{1/4} u^{-1/2} \;.
\ee
Recalling that $h = -i(\tau - \bar{\tau})$, this reproduces the expression obtained in \cite{Belov:2006jd}, equation (6.35), for the norm of the partition function. 

Notice the presence of the factor $u^{-1/2}$ in the norm of the partition function of the self-dual field. This factor depends on the full Hodge star operator acting on $\Omega^{2\ell + 1}(M)$, and not only on its restriction on the intermediate Jacobian. This factor introduces a very non-trivial dependence on the metric of $M$ in the partition function of the self-dual field.

\subsection{Half-torsion}

In this section, we show that the measure $u$, defined in \eqref{EqMesSR}, can be seen as the torsion of a certain complex, in the spirit of the work of Ray and Singer \cite{MR0339293, RaySinger1973, MR0295381}.

In the following discussion, it is useful to recall that the Hodge star operator squares to $(-1)^p$ on the space of $p$-forms on $M$, written $\Omega^p$. Consider the (complexified) De Rahm complex $\Omega^\bullet$ of $M$. Define
\be
\partial = \sum_{p=0}^{2\ell} d|_{\Omega^p} - \sum_{p=2\ell+2}^{4\ell+2} (-1)^{p} d^\dagger|_{\Omega^p} \;, \qquad \partial|_{\Omega^{2\ell+1}} = 0\;,
\ee
where $d$ and $d^\dagger = -\ast d\ast$ are the usual De Rahm differential and codifferential. 
We have $\partial^2 = 0$, so $\partial$ is a differential on the following complex :
\begin{align}
\label{ComplexN}
0 \stackrel{\partial}{\rightarrow} (\Omega^0 \oplus \Omega^{4\ell + 2})_{\rm SD} \stackrel{\partial}{\rightarrow} & \; (\Omega^1 \oplus \Omega^{4\ell + 1})_{\rm SD}  \stackrel{\partial}{\rightarrow} ... \notag \\ ... \stackrel{\partial}{\rightarrow} & \; (\Omega^{2\ell} \oplus \Omega^{2\ell + 2})_{\rm SD} \stackrel{\partial}{\rightarrow} (\Omega^{2\ell + 1})_{\rm SD} \stackrel{\partial}{\rightarrow} 0 \;,
\end{align}
where $(.)_{\rm SD}$ is the projection on the eigenspace of the Hodge operator with eigenvalue $-(-i)^p$. As any element in $(\Omega^p \oplus \Omega^{4\ell + 2-p})_{\rm SD}$ can be written as $\omega -i^p\ast \omega$, $\omega \in \Omega^p$, we can check that $\partial$ preserves the self-duality condition. The complex \eqref{ComplexN} is elliptic and the formal adjoint of $\partial$ is easily computed:
\be
\partial^\dagger = \sum_{p=0}^{2\ell+1} d^\dagger|_{\Omega^p} - \sum_{p=2\ell+1}^{4\ell+2} (-1)^{p} d|_{\Omega^p} \;, \;.
\ee
The associated Laplacian coincides with the usual one: 
\be
\Delta = (\partial + \partial^\dagger)^2 = (d + d^\dagger)^2 \;.
\ee

By analogy with the definitions of Ray and Singer \cite{MR0339293, RaySinger1973, MR0295381}, the torsion of the complex \eqref{ComplexN} is
\be
\label{EqDefSDTorsion}
\ln T_{\rm SD} = \frac{1}{2}  \sum_{p =0}^{2\ell+1} p (-1)^p \ln \det\,\!\!' (\Delta |_{\Omega_{\rm SD}^p(M)}) \;,
\ee
where $\det\,\!\!'(X)$ denotes the zeta-regularized determinant of $X$ on the complement of its kernel. We can rewrite $T_{\rm SD}$ as follows:
\begin{align}
\ln T_{\rm SD} = &\; \frac{1}{2}  \sum_{p =0}^{2\ell+1} p (-1)^p \ln \det\,\!\!' (\Delta |_{\Omega_{\rm SD}^p(M)}) \notag \\ 
= &\; \frac{1}{2} \sum_{p=0}^{2\ell} p (-1)^p \ln \det\,\!\!' (\Delta |_{\Omega^p(M)}) + \frac{1}{2}(-1)^{2\ell+1} (2\ell+1) \ln \det\,\!\!' (d d^\dagger  |_{\Omega^{2\ell+1}(M) \cap {\rm Im} d}) \\ 
= &\; -\frac{1}{2} \sum_{p=0}^{2\ell} (-1)^p \ln \det\,\!\!' (d^\dagger d|_{\Omega^p(M) \cap {\rm Im} d^\dagger}) \;. \notag
\end{align}
To go from the first to the second line, we used for the first term the fact that for each self-dual form with no component on $\Omega^{2\ell+1}(M)$, its projection on the space of forms with degree less than $2\ell+1$ shares the same eigenvalue of the Laplacian. To understand the form of the second term, note that on the complement of the kernel of the Laplacian, there is a bijection $d d^\dagger$ between $\Omega^{2\ell+1}_{\rm SD}(M)$ and the kernel of $d$, with inverse $\Delta^{-1}(1-i^{2\ell+1}\ast)$. This bijection commutes with $\Delta$, hence we can replace the determinant of the Laplacian on the space of self-dual forms with the determinant of $d d^\dagger$ on ${\rm Im} d$. To get the final form, we used the fact that $d d^\dagger|_{\Omega^p(M) \cap {\rm Im} d}$ and $d^\dagger d|_{\Omega^{p-1}(M) \cap {\rm Im} d^\dagger}$ are isospectral.

Comparing with \eqref{EqMesSR}, we see that 
\be
u(g) = \prod_{p = 0}^{2\ell} \left ( V_p^{-1} \right ) (T_{SD})^{-1} \;,
\ee
where $V_p$ is the volume of the torus of harmonic $p$-form in the $L^2$ metric. 

The torsion $T_{SD}$, known as Cheeger's half-torsion, is the object encoding the non-trivial dependence of the partition function of the self-dual field on the metric of the underlying manifold. It was defined in an unpublished work by Cheeger in the 80's. Although it is a close cousin of Ray-Singer analytic torsion, it has a highly non-trivial metric dependence. To our knowledge, it appeared explicitly only once in the literature, in a review by Branson \cite{Branson2005}\footnote{This paper can be downloaded at the following address: \href{http://www.dml.cz/dmlcz/701742}{http://www.dml.cz/dmlcz/701742}.}. In this paper, it was shown that its variations under conformal variation of the metric could be expressed as an integral of a certain density on the manifold. The explicit form of the density was however not completely determined. 

A variational formula for the half-torsion can be derived \cite{Monnier}, in the spirit of the work of Bismut and Lott \cite{1995} on Ray-Singer torsion. However it involves in its current form the asymptotic expansion of a certain heat kernel that is hard to compute explicitly.

\section{Index theory}

\label{SecIndex}

\subsection{Determinant bundle}

Given an elliptic complex like \eqref{ComplexN}, its associated torsion can be used to define a natural Hermitian structure on its determinant bundle, the Quillen metric \cite{0603.32016} (see also \cite{Freed:1986hv}). The function $u$, which determines the Hermitian structure on the quantum bundle $\mathscr{H}^\eta$, can therefore be thought of as the square root of the Quillen metric on the determinant bundle $\mathscr{D}^\eta$ of the complex \eqref{ComplexN}. This strongly suggests that $\mathscr{H}^\eta$ should be a square root of $\mathscr{D}$. Let us see that it is indeed the case, modulo torsion.

$\mathscr{D}$ is defined as follows:
\be
\mathscr{D} = \bigotimes_{p = 0}^{2\ell} {\rm Det}(\mathcal{H}^p \oplus \mathcal{H}^{4\ell+2-p})_{\rm SD}^{(-1)^p} \otimes {\rm Det}(\mathcal{H}^{2\ell+1}_{\rm SD})^{-1} \simeq \bigotimes_{p = 0}^{2\ell} {\rm Det}(\mathcal{H}^p)^{(-1)^p} \otimes {\rm Det}(\mathcal{H}^{2\ell+1}_{\rm SD})^{-1} \;,
\ee
where ${\rm Det}(\mathscr{E})$ denotes the determinant line bundle of a vector bundle $\mathscr{E}$, and $\mathcal{H}^\bullet$ is the space of harmonic forms on $M$, seen as a bundle on $\cCn$. The last identity comes from the bundle isomorphism $(1 - i^p\ast): \mathcal{H}^p \rightarrow (\mathcal{H}^p \oplus \mathcal{H}^{4\ell+2-p})_{\rm SD}$.

The line bundles ${\rm Det}(\mathcal{H}^p)$, $p = 0,...,2\ell$ are trivial, so they cannot contribute to the integral Chern class of $\mathcal{D}$. Indeed, $\mathcal{H}^p$ is the complexification of $\mathcal{H}^p_{\mathbbm{R}}$, the bundle of real harmonic $p$-forms. Then, ${\rm Det}(\mathcal{H}^p) \simeq {\rm Det}(\mathcal{H}_{\mathbbm{R}}^p)^{\otimes 2}$ implies that ${\rm Det}(\mathcal{H}^p)$ is a trivial bundle, as the characteristic class of ${\rm Det}(\mathcal{H}_{\mathbbm{R}}^p)$ is necessarily $\mathbbm{Z}_2$-valued. 

Therefore $\mathscr{D} \simeq {\rm Det}(\mathcal{H}^{2\ell+1}_{\rm SD})^{-1}$, which is nothing but the bundle $\mathscr{K}$ introduced in section \ref{SecAnom}. This shows that, at least modulo torsion, $\mathscr{H}^\eta$ is a square root of $\mathscr{D}$.

\subsection{Dirac operator and local anomaly}

\label{SecDOpLocAn}

We are now ready to make contact with the result of Alvarez-Gaum\'e and Witten \cite{AlvarezGaume:1983ig} on the local gravitational anomaly of the self-dual field.

Let us define the collapsed complex associated to \eqref{ComplexN}:
\be
\label{EqCollapsComplex}
D := \partial + \partial^\dagger = d + d^\dagger : (\Omega^{\rm even})_{\rm SD} \rightarrow (\Omega^{\rm odd})_{\rm SD} \;.
\ee
Let $\mathscr{S}$ be the spin bundle of $M$, and $\mathscr{S}_+$ and $\mathscr{S}_-$ its components with respect to the $\mathbbm{Z}_2$-grading . We have isomorphisms $(\Omega^{\rm even})_{\rm SD} \simeq \mathscr{S}_+ \otimes \mathscr{S}_+$, $(\Omega^{\rm odd})_{\rm SD} \simeq \mathscr{S}_- \otimes \mathscr{S}_+$. \footnote{It is clear from the definition of the Dirac operator \eqref{EqCollapsComplex} that it does not requires 
$M$ to be spin. If it is not, the isomorphisms are valid locally, over open patches in $M$.} We see that $D$ is the Dirac operator on $M$ coupled to chiral spinors. Moreover, the determinant bundle of the index bundle of $D$ is isomorphic to $\mathscr{D}$. This is due to the fact that the kernel and the cokernel of $D$ are given by spaces of harmonic forms, whose dimensions are constant over the space of metrics. The non-triviality of the index bundle is therefore completely contained in the determinant of the space of zero modes (see section 4 of \cite{springerlink:10.1007/BF01458075}). 

$D$ is exactly the Dirac operator used by Alvarez-Gaum\'e and Witten to compute the local anomaly of the self-dual field \cite{AlvarezGaume:1983ig,AlvarezGaume:1984dr}. Their idea was to supplement the anomalous self-dual field with auxiliary non-anomalous $p$-forms, $p = 0,..., 2\ell$ to obtain a Clifford representation and use index theory. For a family of manifolds with metric $Y \rightarrow B$ with fiber $M$, a straightforward application of the Bismut-Freed formula \cite{MR853982, MR861886, Freed:1986hv} yields the usual formula for curvature of $\mathscr{D}$ 
\be
\label{EqCurvDetBd}
R(\mathscr{D}) = 2\pi i \left(\int_M \frac{1}{4}L(R(TM))\right)^{(2)} \;,
\ee
where $R(TM)$ is the curvature of $TM$, seen as a bundle over $M \times B$. The exponent ${(2)}$ denotes the projection on the two-form component and the L-genus is defined by
\be
L(R) = 2^{2\ell +2} {\rm det}^{1/2} \frac{R/4\pi}{\tanh{R}/4\pi} \;.
\ee 
It is interesting to note that only $\mathscr{D} \simeq (\mathscr{H}^\eta)^2$ has an analytic interpretation as the determinant bundle of a Dirac operator. In Lorentzian signature, a symplectic Majorana condition can be imposed on the complex \eqref{EqCollapsComplex}, which makes it clear that 
$\mathscr{D}$ is the square of a line bundle. In Euclidean signature, the formula \eqref{EqCurvDetBd} is usually inconspicuously divided by two, yet there is no warranty that the resulting two-form defines an integral cohomology class and one can question the existence of the anomaly bundle of the self-dual field on manifolds of Euclidean signature. In dimension $8\ell + 2$, the fact that the spinors are quaternionic in dimension $8\ell + 4$ implies that the index of $D$ is even and that $R(\mathscr{D})$ defines an even cohomology class \cite{Freed:1986hv}. However, no similar argument is available in dimension $8\ell + 6$, the case of interest for the five-branes. 


This puzzle should be resolved by the investigation of the anomaly bundle with the formalism of Hopkins and Singer \cite{hopkins-2005-70}. For instance in their work, the formula \eqref{EqCurvDetBd} is ``divided by two'' in a non-trivial way involving a quadratic refinement of the intersection form, a choice of which is equivalent to a choice of characteristic. We also believe the global anomaly can be determined by evaluating their action on mapping tori. We plan to come back to these issue in a future publication.

%

\subsection*{Acknowledgments}

I would like to thank Jean-Michel Bismut, Emanuel Diaconescu, Rod Gover, Matthew Gursky, Greg Moore, Daniel Persson and Boris Pioline for discussions and correspondence. Thanks also goes to the community at MathOverflow.net, especially Greg Kuperberg, Tim Perutz and Andrew Putman. Most of this work was done while I was visiting the New High Energy Theory Center at the Physics department of Rutgers University, that I would like to thank for hospitality and generous financial support. This work was supported in part by grant PBGE2--121187 of the Swiss National Science Foundation and by a Marie Curie intra-European fellowship, grant agreement number 254456.

\appendix

\section[The determinant of the (2,0) supermultiplet on a Calabi-Yau]{The determinant of the (2,0) supermultiplet on a Calabi-Yau threefold\footnote{The work pertaining to this section was started upon the reading of a draft of the paper \cite{Alexandrov:2010ca}. I would like to thank the authors for sharing it with me.}
}

\label{SecOne-Loop}

The self-dual field in six dimensions appears notably in the (2,0) supermultiplet that lives on the worldvolume of the M5- and NS5-branes in M-theory and type IIA string theory. The one loop determinant of this supermultiplet governs the amplitude of the corrections to low-energy supergravity by five-brane instantons. We want to compute its norm, in the case when $M$ is a Calabi-Yau threefold. We focus only on the contribution from the non-zero modes.

The worldvolume of a single five-brane contains scalar and fermionic fields in addition to a real self-dual field. There are five real scalars, corresponding for the M5-brane to the five transverse directions to the brane. In the case of the type IIA NS5-brane, one of the scalars is an axion associated to the M-theory circle. The fermions can be described as follows. Let $S^+_N$ be the chiral spinor bundle associated to the normal bundle of the worldvolume of the five-brane. The structure group of the latter is $SO(5) \simeq USp(4)$, so a symplectic Majorana condition can be imposed on $\mathscr{S}_N$, yielding a spinor bundle $\mathscr{S}^{1/2}_N$ satisfying $\mathscr{S}^{1/2}_N \oplus \mathscr{S}^{1/2}_N = \mathscr{S}^+_N$. If we denote by $\mathscr{S}^+_T$ the chiral spinor bundle of the worldvolume $M$ of the five-brane, the fermions are sections of $\mathscr{S}^+_T \otimes \mathscr{S}^{1/2}_N$. $\mathscr{S}^+_T$ is four-dimensional, while $\mathscr{S}^{1/2}_N$ is two dimensional. The eight fermionic degrees of freedom match the $3 + 5$ bosonic degrees of freedom coming from the self-dual field and the scalars \cite{Schwarz:1997ed}.

We consider now a space-time of the form $M \times \mathbbm{R}^4$, where $M$ is a Calabi-Yau threefold and we let a five-brane wrap $M$ at the origin of $\mathbbm{R}^4$. Let us denote by ${\rm det}' \Delta_{pq}$ the zeta-regularized determinant of the Laplacian on $M$ acting on the complement of its kernel in the space of $(p,q)$-forms. On a Calabi-Yau, these determinants satisfy the relations (see for instance \cite{Pestun:2005rp})
\begin{align}
& \; {\rm det}' \Delta_{00} = {\rm det}' \Delta_{33} = {\rm det}' \Delta_{30} = {\rm det}' \Delta_{03} \;, \notag\\
& \; {\rm det}' \Delta_{10} = {\rm det}' \Delta_{01} = {\rm det}' \Delta_{32} ={\rm det}' \Delta_{23} \\
& \; = {\rm det}' \Delta_{20} = {\rm det}' \Delta_{02} = {\rm det}' \Delta_{13} = {\rm det}' \Delta_{31} \;, \notag\\
& \; {\rm det}' \Delta_{11} = {\rm det}' \Delta_{21} = {\rm det}' \Delta_{12} = {\rm det}' \Delta_{22} \;. \notag
\end{align}

We already saw that the norm of the one-loop determinant of the self-dual field on the worldvolume of the five-brane is given by the square root of the Cheeger half-torsion of $M$. Using \eqref{EqDefSDTorsion}, we see that it can be written in terms of the determinants above as:
\be
(T_{SD})^{1/2} = ({\rm det}' \Delta_{00})^{-3/4} ({\rm det}' \Delta_{10})^{1/2} ({\rm det}' \Delta_{11})^{-1/4}\;.
\ee
 
Now $({\rm det}' \Delta_{00})$ is simply the determinant of the Laplacian on the space of functions on $M$, so the scalars give an extra factor $({\rm det}' \Delta_{00})^{-5/2}$.

The contribution of the fermions is more subtle to compute. First recall that a Calabi-Yau manifold admits a covariantly constant chiral spinor $\psi_0 \in \Gamma(\mathscr{S}^+_T,M)$, that we can see as a section of a line bundle $\mathscr{P}$ over $M$. The complement of $\mathscr{P}$ in $\mathscr{S}^+_T$ is a bundle $\tilde{\mathscr{S}}$ with $SU(3)$ structure group and the Laplacian preserves the decomposition $\mathscr{S}_T = \tilde{\mathscr{S}} \oplus \mathscr{P}$. Note also that in our setup, the normal bundle of $M$ is trivial, so $\mathscr{S}^{1/2}_N$ is a trivial bundle as well, with fibers of dimension 2. 


Let us write the generators of the Clifford algebra of the cotangent space of $M$ as $\gamma_\mu$. We have a relation between the holomorphic 3-form $\Omega$ on the Calabi-Yau and the covariantly constant spinor: $\Omega_{\mu\nu\sigma} = \psi_0^T \gamma_{\mu\nu\sigma} \psi_0$, $\gamma_{\mu\nu\sigma}$ being the totally antisymmetric product of three gamma matrices (see for instance \cite{citeulike:1923911}, page 378 and following). We deduce that, if we denote respectively holomorphic and anti-holomorphic indices by underlining and overlining, the covariantly constant spinor satisfies $\gamma^{\overline{\mu}} \psi_0 = 0$. Now we can get an explicit form of the isomorphism between $\mathscr{S}^+_T \otimes (\mathscr{S}^+_T)^\ast$ and $\bigwedge T^\ast M$ by considering the transformation properties of forms and spinors under the $SU(3)$ holonomy group of the Calabi-Yau:
\be
\gamma^{\underline{\mu}_1}...\gamma^{\underline{\mu}_p} \psi_0 \otimes (\gamma^{\underline{\nu}_1}...\gamma^{\underline{\nu}_q} \psi_0)^\ast \rightarrow dz^{\mu_1} \wedge ... dz^{\mu_p} \wedge d\bar{z}^{\nu_1} \wedge ... d\bar{z}^{\nu_q}\;.
\ee
Therefore we find that $\tilde{\mathscr{S}} \otimes \mathscr{P}^\ast \simeq \Omega^{(2,0)}(M)$ and $\mathscr{P} \otimes \mathscr{P}^\ast \simeq \Omega^{(0,0)}(M)$. 

By tensoring sections of $\tilde{\mathscr{S}}$ with $\psi_0^\ast$, we get a bijection
\be
\label{EqIsomSpinFormCY}
\tilde{\mathscr{S}} \oplus \mathscr{P} \rightarrow \Omega^{(2,0)}(M) \oplus \Omega^{(0,0)}(M)
\ee
Because $\psi_0$ is covariantly constant, the map \eqref{EqIsomSpinFormCY} commutes with the Laplacian and the contribution of the fermions to the one-loop determinant is given by $ {\rm det}' \Delta_{20} \, {\rm det}' \Delta_{00} =  {\rm det}' \Delta_{10} \, {\rm det}' \Delta_{00}$.


Assembling all the contributions to the one-loop determinant of the (2,0) supermultiplet, we get:
\be
(T_{BCOV})^{-1/2} = ({\rm det}' \Delta_{00})^{-9/4}({\rm det}' \Delta_{10})^{3/2} ({\rm det}' \Delta_{11})^{-1/4} \;.
\ee
$T_{BCOV}$ is the Bershadsky-Cecotti-Ooguri-Vafa torsion, that appears in the one-loop determinant of the B-model. Such a relation between the determinants of the partition functions of the topological string and the five-brane has been conjectured in \cite{Dijkgraaf:2002ac, Alexandrov:2010ca}. Note as well that the quantization of the intermediate Jacobian of a Calabi-Yau threefold also appears naturally in relation to the B-model \cite{Witten:1992qy, Gunaydin:2006bz, Aganagic:2006wq, Schwarz:2006br}. However to our knowledge the one-loop determinant has not been derived in this framework.

{
\small

\providecommand{\href}[2]{#2}\begingroup\raggedright\endgroup

}

\end{document}